\documentclass[aps,pre,showkeys,groupedaddress,amsmath,amssymb,floatfix,twocolumn]{revtex4-1}
\usepackage{times}

\usepackage{lipsum}
\usepackage{dsfont,eucal,mathrsfs}
\usepackage{bm}
\usepackage{verbatim}
\usepackage{color} 
\usepackage{graphicx}
\newcommand{\bo}{\boldsymbol}

\renewcommand{\mathbf}{\boldsymbol}
\renewcommand{\mathcal}{\mathscr}

\newtheorem{theorem}{Theorem}

\usepackage{ragged2e}
\usepackage{hyperref}
\def\kapppa{k}

\begin{document} 
%----------------------------- HEADER ----------------------------------
\title{Revealing dynamics, communities and criticality from data}
\author{Deniz Eroglu$^{1,2,3}$}
\email{deniz.eroglu@khas.edu.tr}
\author{Matteo Tanzi$^{2,4}$}
\author{Sebastian van Strien$^{2}$}
\author{Tiago Pereira$^{1,2}$}
%\email{yamir.moreno@gmail.com}
\affiliation{$^{1}$Instituto de Ci\^encias Matem\'aticas e Computa\c{c}\~ao, Universidade de S\~ao Paulo, S\~ao Carlos, Brazil}
\affiliation{$^{2}$Department of Mathematics, Imperial College London, London, UK}
\affiliation{$^{3}$ Department of Bioinformatics and Genetics, Kadir Has University, 34083 Istanbul, Turkey}
\affiliation{$^{4}$Courant Institute of Mathematical Sciences, New York University, New York NY, USA}
%-----------------------------------------------------------------------

\begin{abstract}
  Complex systems such as ecological communities and neuron networks are essential parts of our everyday lives.  These systems are composed of units which interact through intricate networks. The ability to predict sudden changes in the dynamics of these networks, known as critical transitions, from data is important to avert disastrous consequences of major disruptions. Predicting such changes is a major challenge as it requires forecasting the behaviour for parameter ranges for which no data on the system is available. We address this issue for  networks with weak individual interactions and chaotic local dynamics. We do this by building a model network, termed an {\em effective network}, consisting of the underlying local dynamics and a statistical description of their interactions. We show that behaviour of such networks can be decomposed in terms of an emergent deterministic component and a {\em fluctuation} term. Traditionally, such fluctuations are filtered out. However, as we  show,  they are key to accessing the interaction structure. 
 {
  We illustrate this approach on synthetic time-series of  realistic neuronal interaction networks of the cat cerebral cortex and on experimental multivariate data of optoelectronic oscillators.  
 }
  We reconstruct the community structure by analysing the stochastic fluctuations generated by the network  and predict critical transitions for coupling parameters outside the observed range.
\end{abstract}

\maketitle 

%\tableofcontents

\section{Introduction}

We are surrounded by a range of complex networks composed of many units forming an intricate network of interactions. Neuron networks form an important class of examples where the interaction structure is heterogeneous \cite{kandel2000}. Because changes in the interaction can have massive ramifications on the system as a whole, it is desirable to predict such disturbances and thus enact precautionary measures to avert potential disasters.  For instance, neurological disorders such as Parkinson's disease, schizophrenia, and epilepsy, are thought to be associated with an anomalous interaction structure among neurons \cite{bohland2009}. As in the case of neuron networks, it is impossible to directly determine the interaction structure.  Therefore, a major scientific challenge is to develop techniques  using measurements of the time evolution of the nodes  to indirectly recover the network structure and predict the network behaviour when the interactions change. 
 
The literature on data-based network reconstruction is vast. Reconstruction methods can be classified into {\it model-free} methods and \emph{model-based} methods.  The former identify the presence and strength of a connection between two nodes  by measuring the {\it dependence} between their time-series in terms of:  correlations \cite{de2004discovery,reverter2008combining}, mutual information \cite{butte1999mutual}, maximum entropy distributions \cite{braunstein2008inference,cocco2009neuronal}, Granger causality, and causation entropy  \cite{bressler2011wiener,ladroue2009beyond}. Such methods alone do not provide information on the dynamics, which is necessary to predict critical transitions. Model-based methods provide estimates (or assume a priori knowledge) of the dynamics and interactions, and use this knowledge to reconstruct the network structure. When the interactions are strong, the network structure can be recovered \cite{casadiego2017,han2015,wang2016}.  For a more extensive account of reconstruction (model-free and -based) methods see the reviews \cite{wang2016,nitzan2017revealing,stankovski2017}. 

In many applications, the behaviour of isolated nodes is chaotic and the interaction is weak \cite{schneidman2006,haas2015,kandel2000}. The network structure typically has communities and hierarchical organisations such as the rich-clubs \cite{heuvel2011}.  As the interaction strength per connection is weak and the statistical  behaviour of the nodes is persistent,  the influence of each node on the network corresponds essentially to a random signal.
%Furthermore, because of the chaotic dynamics and weak coupling,. Through the interactions of the network these signals  determine the network dynamics.
%that is superposed to the randomness generated by the chaos in the local dynamics. 
Existing  techniques fail to reconstruct a model from the data, as they require the interaction to be of the same magnitude as the isolated dynamics. 
%
% \textcolor{red}{
 In our setting,  only the cumulative contribution of many links matter and the network signals decompose into a  deterministic and a fluctuation term. The  latter, which is usually filtered out, turns out to give crucial information on the network structure and is fundamental to our approach.
 %} 

In this paper, we introduce the notion of an {\em effective network} which aims to model a complex system from observations of the nodes evolution when the network has a heterogeneous structure,  the strength of interaction is small and  local dynamics are highly erratic. This approach starts by reconstructing the local dynamics from observations of nodes with relatively few connections, and then recover the interaction function from observations of the highly connected nodes whose dynamics are the most affected by the interactions as a result of the multitude of connections they receive from the rest of the network \cite{park2013,pereira2017}. A key achievement is that this reconstruction enables us to identify  community structures 
also when the coupling is only weak. 
Moreover,  it recovers enough information to forecast  and  anticipate the network behaviour,  even in situations where the parameters of the system change into ranges that have not been previously encountered.

\subsection{Complex networks of nonlinear systems} 

We consider  networks with $N$ nodes with chaotic isolated dynamics and pairwise interactions. The network is described by its adjacency matrix  $\bo A$, whose entry $A_{ij}$ equals $1$ if node $i$ receives a connection from $j$ and equals $0$ otherwise. The  time evolution of the state $\bo x_i(t)$ of node $i$ at time $t$ is expressed  as
\begin{eqnarray}
\bm{x}_i(t+1)&=&\bm{F}_i (\bm{x}_i(t))+\alpha \sum_{j=1}^{N} A_{ij} \bm{H}(\bm{x}_i(t),\bm{x}_j(t)). 
\label{eq1}
\end{eqnarray} 
When performing reconstruction, the isolated local dynamics $\bm{F}_i\colon M\to M$, the coupling function $\bm{H}$, the coupling parameter $\alpha$ (that is small), the adjacency matrix $\bo A$,  and even the dimension $k$ is the degree of the space $M$, are assumed to be  {\em unknown}. These equations model  important complex systems such as neuron networks \cite{izhikevich2007},  smart grids\cite{yadav2017,dorfler2013},  superconductors \cite{watanabe1994}, and  cardiac pacemaker cells \cite{winfree2001}.  

\subsection{Main assumptions}

Our three assumptions are: $(a)$ the local dynamics are close to some unknown {\em ergodic} and {\em chaotic} map $\bm{F}$ (that is, $\| \bm{F} - \bm{F}_i \| \le \delta$, which is often the case in applications \cite{pinto2000,eroglu2017}). $(b)$ The network connectivity is {\em heterogeneous}, which means that the  number of incoming connections at a node  $i$ (given by its degree $k_i = \sum_j A_{ij}$) varies widely across the network.  $k_i$ is large
 for a few nodes called {\em hubs}. $(c)$ $\alpha$ is such that, denoting by $\Delta=\max_i k_i$ the maximum number of connections  $\alpha \Delta$ is of the same magnitude of $\bm{DF}$.  Assumptions $(a)$ and $(c)$ imply that only the cumulative effect of the coupling is important.  A prime example  is the cat cerebral cortex which possesses inter-connected regions split into communities with a hierarchical organization {  as well as modular and disassortative rich-clubs}. This network has heterogeneous connectivity, chaotic motion and weak coupling \cite{scannell1993,scannell1995,zamora-lopez2010}. Other examples include the drosophila optic lobe network \cite{takemura2013,garcia-perez2018}. For a given dataset, our effective network first tests whether the underlying system satisfies assumptions $(a)-(c)$, and, if so, reconstructs the model.

We assume the availability of  a time series of  observations 
\[y_i(t)=\phi( {\bo x}_i(t))\]  where $\phi$ is a projection to a variable on which unit interactions depend. This situation occurs frequently in applications; as with  measurements of membrane potentials in neurons. 
 
\section{Effective networks recover  structure and dynamics}

To obtain an {\em effective (reconstruction of the) network} from observations, we combine statistical analysis, machine-learning techniques, and dynamical systems theory for networks. An effective network provides:  local evolution laws and averaged interactions for each unit that, in combination, closely approximate  the unit dynamics; a network with the same degree distribution and community structures as the original system. We use the term ``effective" because it gathers sufficient data to reproduce the behaviour of the original network and predict its critical transitions.

Using  our assumptions for the network and local dynamics, we can show that the evolution at each node will have low-dimensional excursions over finite time scales. More precisely, the evolution rule at node $i$ is given by
\begin{eqnarray}
\bo G_i(\bo x_i)  = \bo F_i(\bo x_i) + \beta_i \bo V(\bo x_i(t)) \nonumber
\end{eqnarray}
where $\bo F_i \approx \bo F$  is the isolated dynamics,  
\[ \beta_i = \alpha k_i\] is the rescaled degree, and 
\[\bo V (
\bo{x}) = \int \bo{H}(\bo{x},\bo{y}) d  \mu (\bo{y})
\] 
where $\mu$ is physical measure of the isolated dynamics. $\bm{V}$  takes into account the cumulative effect of interactions on node $i$.  The true dynamics   
\begin{eqnarray}
\bo x_i(t+1) = \bo G_i(\bo x_i(t)) + \bo \xi_i(t) \nonumber
\end{eqnarray}
is influenced by a fluctuation term $\bo \xi_i(t)$ that is  small for an interval of time which is exponentially large and depends on the state of neighbours of the $i$th node.
 This low-dimensional reduction  has been rigorously established in test cases (see \cite{pereira2017}). See Appendix  for further information. 

The approximation described above applies to the measured state variable $y_i(t)$. {First of all we pre-process the data according to the system under study} (see  Appendix \ref{App:addsim}). The processed variable is still referred to as $y_i(t)$. Takens reconstruction tells us that  $y_i(t+1)$  is a nonlinear function of $k+1$ past points $y_i(t), \dots y_i(t-k)$, for a given number $k$ provided by the approach. Here, we focus on the case when $k=1$, which occurs in many real-world examples, and discuss cases with $k\ge2$ in Appendix \ref{App:addsim}. This means that 
\begin{eqnarray} \label{eq:meanfield}
y_i(t+1) = g_i(y(t))+ \xi_i(t)
\end{eqnarray}
where $g_i=f_i(y_i(t))+\beta_i v(y_i(t))$, and $v$ is the corresponding projection of effective coupling $\bo{V}.$
%\begin{eqnarray}
%y_i(t+1)=f_i(y_i(t))+\beta_i v(y_i(t))+\xi_i(t) \mbox{ where } \beta_i = \alpha k_i   %\nonumber
%\end{eqnarray}
% and coarsely classify the nodes by their degrees. 
%For instance, in rich-club motifs, the network has low-degree nodes organised in communities and high-degree in the rich-club. 
%Since $\beta_i=\alpha k_i$ is small for low-degree nodes $i$,  the evolution rules at such nodes will be similar to the isolated dynamics. Thus we identify 
%which nodes are low-degree nodes and we can recover the local dynamics $f$.  Next, we use Eq. (\ref{eq:meanfield}) and a classification
%of nodes by their time-series, to obtain the coupling function and estimate the degree distribution $k_i$ and the coupling parameter $\alpha$. 
\subsection{Reconstruction procedure}
An effective network is obtained in three main steps: 

{
{\bf Step 1: Reduced dynamics.}  
We employ Takens reconstruction. If the time series is high dimensional, we discard it. Otherwise, once we are in the appropriate dimension, we estimate and learn the rule $g_i$. We decompose  $g_i$ as a linear combination of basis functions, tailored to the application. The parameters of the basis functions are obtained by performing a $10$-fold cross-validation with $90\%$ training and $10\%$ test \cite{shandilya2011, james2013} (see Appendix \ref{app:en}). As the dynamics is low-dimensional other techniques such as compressive sensing \cite{brunton2015,wang2011} or embedding  \cite{judd1998} can be also employed.}

 {
 {\bf Step 2: Isolated dynamics and effective coupling.} We run a model-free estimation that coarsely classify nodes according to their degree by  assigning to every pair of $y_i$ and  $y_j$  a \emph{Pearson distance} $s_{ij}\ge0$  such that $s_{ij} \approx 0$ if the attractors of $i$ and $j$ are similar  and $s_{ij} \approx 1$ if they are distinguishable. The higher the number of nodes with behaviour different from $i$, the larger  the intensity  $S_{i} = \sum_{j}s_{ij}$. Low degree nodes have typically small $S_i$ while for hubs this quantity is large.  Notice that for the low-degree nodes, $\alpha \kapppa_i v$ is negligible and the dynamics at the low-degree nodes are  close to $f$. Therefore we use $g_i$  at the identified low degree nodes to obtain an approximation for $f\approx g_i$, while $g_i$  at hub nodes allows to estimate $\beta_i v\approx g_i- f$.  We estimate $\beta_i$ by Bayesian inference.}

 {
{\bf Step 3: Network structure and communities.} Since $\beta_i=\alpha k_i$, we can recover the network's degree distribution from $\beta_i$.
Then, having the local rules $g_i$, we can decompose the time series in terms of a low-dimensional deterministic part and the fluctuation term $\xi_i$, and use this last term to recover community structures. If nodes $i$ and $j$ interact with the same nodes, they are subject to the same inputs and  the correlation Cor$(\xi_j,\xi_i)$ is high. If not,  Cor$(\xi_i,\xi_j)$ is nearly zero due to the decay of correlations in the deterministic part. Thus, Cor$(\xi_j,\xi_i)$ is high when nodes $i$ and $j$ have high {\em matching index} (high fraction of common connections), and are likely to belong to the same cluster. Given the matrix $\rho_{ij}=\mbox{Cor}(\xi_i,\xi_j)$, we estimate the adjacency matrix $\bm{A}$  by thresholding the correlation matrix as $A_{ij}= \Theta(\rho_{ij} > \tau)$, where $\Theta$ is a Heaviside step function and the value of the threshold $\tau$ between 0.3 and 0.6.  We then apply the modularity-based Louvain method \cite{blondel2008} on $\bm{A}$ to detect communities.}

{That  Cor$(\xi_j,\xi_i)$ is high when nodes $i$ and $j$ have high matching index, is true for generic coupling as shown by the following argument. In  general, the coupling function is a sum of terms 
$
h(x,y) = u(x) v(y).
$
 This leads to noise terms 
\begin{eqnarray*}
\xi_i (t) = u(x_i) \left( \frac{1}{\Delta}\sum_{j} A_{ij} v(y_j) - k_i \int v(y) d\mu(y) \right)
\end{eqnarray*}
where $\mu$ is the physical measure of the local dynamics. Given $i$ and $j$ the sum can be split into common connections to $i$ and $j$ and to the independent connections:
\begin{eqnarray*}
\xi_i  = u(x_i)[\zeta_i(t) + w(t)]  \mbox{~ and ~} \xi_j = u(x_j)[\zeta_j(t) + w(t)]
\end{eqnarray*}
where $w$ is the noise due to the common connections (notice that $w$ has zero mean), and $\zeta_i$, $\zeta_j$ depend on different coordinates and can be assumed to be uncorrelated. Omitting the time index $t$,  the covariance of $\xi_i$ and $\xi_j$ is 
\[
\mbox{Cov}(\xi_i, \xi_j) \approx \mathbb{E}  [(u(x_i) w)( u(x_j)  w) ].
\]
 After some manipulations, we obtain 
\begin{eqnarray}
\mbox{Cov}(\xi_i, \xi_j) &\approx& \langle u \rangle^2 \mbox{Var} (w) 
\end{eqnarray}
so, if 
$
\int u (x) d\mu(x) = 0, 
$ the correlation between the noise will vanish even though they have a common term. Thus, the above scheme is able to recover communities if  $ \langle v \rangle \not=0$. If this condition is not met, the network reconstruction via the $g_i$'s is not possible. We remark that $\langle v \rangle=0$ is a  special condition on the coupling that is destroyed by small perturbations. }

 {  It is crucial that the correlation analysis is restricted to fluctuations $\xi_i$. Since the variance of the deterministic part of $y_{i}$ is larger than that of the small fluctuations $\xi_i$, performing a direct correlation analysis between $y_{i}$ and $y_{j}$ hides all the contributions coming from the covariance between $\xi_i$ and $\xi_j$. Consequently, the correlation of the deterministic part  is close to zero due to the chaotic dynamics,
 {
  as shown in Appendix A.
  }
}

\subsection{Benchmark model for the isolated dynamics}  

We present the effective network methodology applied to networks of neurons. We use synthetic time-series where each  neuron is simulated using the Rulkov model, which has two variables, $u$ and $w$, evolving at different time scales  {as described by $\bm{F}(\bm{x}) = (F_1(u,w),F_2(u,w))$ with 
\begin{eqnarray*}
F_1(u,w) = \frac{\beta}{1+u^2} + w
\quad\mbox{ and }\quad
F_2(u,w) = w - \nu u - \sigma.
\end{eqnarray*}
 The fast variable $u$ describes the membrane potential and  is the state variable measured by the observed time series $y_i(t)$, while $w$ describes the slow currents.  Different combinations of parameters  $\sigma$  and $\beta$  give rise to different dynamical states of the neuron, such as resting, tonic spiking, and chaotic bursts.  To test our procedure we considered two cases:  $\sigma=\nu=0.001$ and $\beta=5.9$, which correspond to tonic spiking, and $\beta=4.4$ which correspond to bursting. As for the coupling, we consider \emph{chemical synaptic coupling}, that is, $ \bm{H}(\bm{x}_i, \bm{x}_j) = (h(u_i,u_j),0)$ with  $h(u_i,u_j) = (u_i - V_s) \Gamma (u_j)$, where 
\begin{eqnarray*}
\Gamma(u_j) = 1/ (1+\exp\{ \lambda( u_j - \Theta_s )\} ),
\end{eqnarray*}
and \emph{electrical synaptic coupling}, $ \bm{H}(\bm{x}_i, \bm{x}_j) = (h(u_i,u_j),0)$ with  $h(u_i,u_j) = u_j - u_i$. 
In the chemical coupling, $V_s$  is a parameter called reverse potential. Choosing $V_s > u_i(t)$, the synaptic connection is excitatory. We take $V_s = 20$, $\Theta_s = - 0.25$, and $\lambda=10$.
} 
   {In addition to Rulkov maps,  we show in the Appendix \ref{App:addsim}  that the approach performs well on a wide range of nonlinear local dynamics such as: doubling maps, logistic maps, Spiking Neurons, Henon maps. We also provide performance analysis for R\"ossler oscillators in Section III of the Supplementary Material. }
%for bursting dynamics. We also validated our methods for 

%\subsection{Models for the heterogeneous network} 
%
%We apply our methodology   {to a wide range of simulated network structures such as scale-free networks with various parameters and rich-club networks, and real world networks such as the Drosophila and the cat cerebral cortex.}   The cat cortex is characterized by a rich-club motif and together with the networks with the Erd\"os-Renyi clusters is used as benchmark to test the performance of our methodology in recovering communities. We use scale-free networks and the optic lobe of D. Melanogaster  to test the ability of our approach in recovering the degree distribution.
% (also in the Supplementary materials). 

\section{Revealing community structure: the rich-club motif}

%\tpa{}

\begin{figure*}[t!]
\centering
\includegraphics[width=1.0\textwidth]{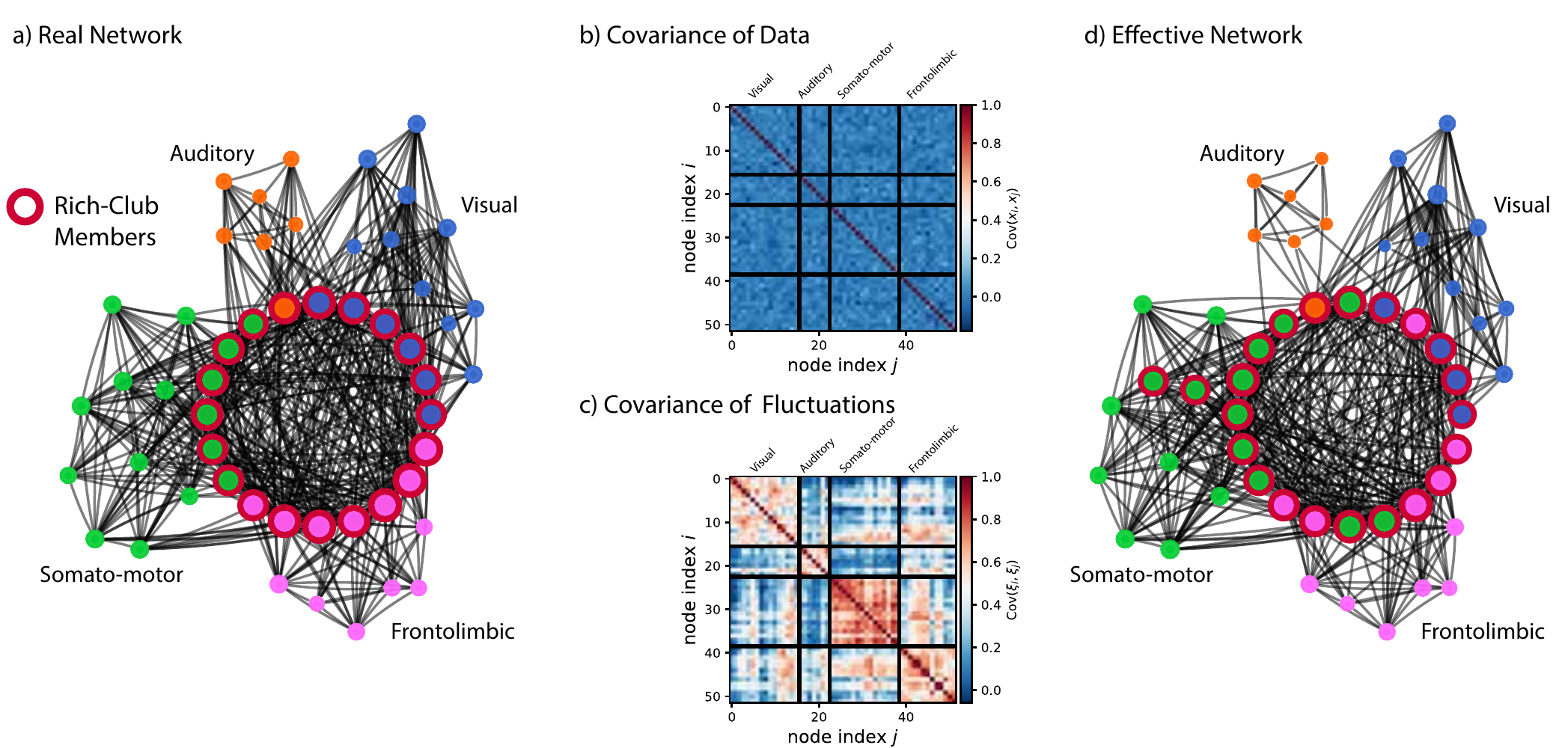}
\caption{{\bf Effective network of the cat cerebral cortex.}  We use the local dynamics as a spiking neuron coupled via electric synapses. (a) The cat cerebral cortex network with nodes colour coded according to the four functional modules. Rich-club members are indicated by  red encircled nodes. (b) \emph{The covariance matrix of the data} cannot detect communities. (c) \emph{The covariance matrix of the fluctuations} can distinguish clusters. This matrix has entries color coded according to the key on the right with red entries corresponding to couple of nodes sharing a large numbers of nearest neighbours in the network, while blue nodes correspond to couple of nodes that share a  small number of common neighbours.   (d) A model in the cat cortex constructed via the effective network approach. From the matrix in (c) we can recover a representative effective network.  The reconstructed network represents the actual network in (a) with good accuracy.}
\label{RC2} 
\end{figure*}

We focus on the network structure of the cat cerebral cortex \cite{zamora-lopez2010}.   The network contains 53 meso-regions arranged in  four communities that follow functional subdivisions;   visual (16 nodes), auditory (7 nodes), somatomotor (16 nodes) and frontolimbic (14 nodes), as shown in Fig \ref{RC2} (a). Some cortical areas (hubs) form a hidden layer called a \emph{rich-club} and are densely connected to each other and the communities. A set of nodes form a rich-club if their level of connectivity exceeds what would be expected by chance alone. The maximum number of connections in this network is $\Delta = 37$.

 {The regions and their connections were discovered by using datasets from tract-tracing experiments \cite{scannell1993,scannell1995}.} The network obtained is weighted. For simplicity  and to improve the performance  in detecting communities, we turn the network into an undirected simple graph \cite{zamora-lopez2010}.
%
%{\color{red}}
%
We simulate each mesoregion as a neuron interacting via electrical synapses and obtain a multivariate data  $\{ y_1(t), y_2(t), \dots, y_N(t) \}$ for a time $T=5000$.
For simplicity, we will denote $y_i = \{ y_i(t)\}_{t=0}^T$.

\subsection{Comparison with previous approaches}
For comparison, we recover the network using  {two widely employed approaches:  functional networks \cite{bettinardi2017,eguiluz2005,bullmore2009}, and sparse recovery techniques \cite{brunton2015,wang2016}. }
The intuition behind the \emph{functional network approach} is that nodes with similar time series have similar characteristics. The functional network can be constructed by the matrix of similarities between nodes via statistical analysis \cite{zhang2006,greicius2003}. As a measure of similarity, we employ a covariance analysis between the time series. The functional network  {cannot detect communities in this case since the time-series at different nodes are essentially uncorrelated }(Fig. \ref{RC2} (b)). Other similarity measures give no significative improvement. See Appendix \ref{app:func_networks} for the details.

 The key idea in \emph{sparse recovery techniques} is to write the dynamics as a linear combination of  basis functions with unknown  coefficients, and the presence of a link is determined when any coefficient of the corresponding interaction is nonzero. { Thus a link is present if the estimated coefficient corresponding to the link is above a given threshold $\sigma$}.
 
 { We implemented the sparse recovery method to our benchmark model when the strength of each connection 
 is of order $\alpha \approx 0.015.$ Hence we have chosen values of $\sigma$ close to this value. The reconstructed network does not identity the clusters correctly as can be seen by comparing the blue and red markers in  Figure \ref{SR}.
 \begin{figure}[h!]
    \centering           
        \includegraphics[width=1.0\linewidth]{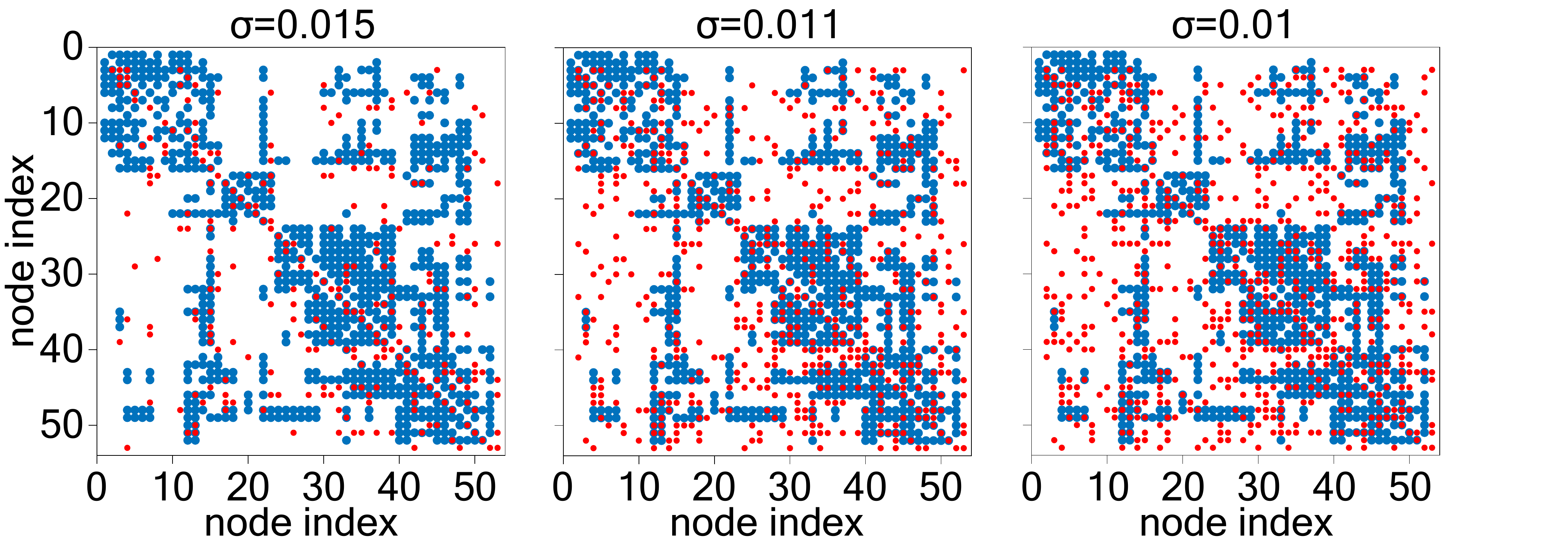}
\caption{{{\bf Sparse recovery method on on cat cerebral cortex.} Sparse recovery is applied to the data  generated by bursting neurons electrically coupled on the cat cerebral cortex. Selecting the threshold parameter $\sigma$ in the method changes the reconstructed network. Here we show the results of sparse recovery method for different enforced sparsity $\sigma$. The nonzero entries of the original network's adjacency matrix are in blue.  The red filled circles represent the nonzero entries in the adjacency matrix of the network reconstructed with the sparse recovery method. As each connection is small in comparison with the isolated dynamics, the sparse recovery tends to neglect them.} }
\label{SR}
\end{figure}
 {In the cases that we are studying here each individual link provides a negligible contribution and only the cumulative effect of many links is relevant. } The coefficients to be recovered are close to zero, and cannot be distinguished from zero terms. A discussion on sparse recovery can be found in Section I of Supplementary Materials. }

\subsection{Community structure via effective networks} 
Remarkably, the effective network is able to recover the community structures (Fig. \ref{RC2} (c)). 
Using Step 1, 2, and 3 we obtain a model for the isolated dynamics, coupling function, distribution of degrees, and correlations Cor$(\xi_i,\xi_j)$.  To apply the method of community detection in \cite{blondel2008}, we threshold the matrix of correlations, Fig. \ref{RC2} (c),  considering nodes $i$ and $j$ linked  only when the correlations were greater than $0.5$. We test threshold values ranging from  $0.3$ to $0.6$ and obtained the similar results as the distribution of the entries of the matrix of correlations is unimodal and has a peak near 0.5.   {We use the algorithm in \cite{colizza2006} to compute the rich-club coefficients for each node. The coefficient depends on the degree and is a number between 0 and 1. We assigned to the rich-club the nodes with coefficient at least 0.8.}  As shown by Figure \ref{RC2} (d), the effective network methodology is able to classify  the nodes in the network according to their function.

 {Notice that our model predicts the presence of a link between two nodes $i$ and $j$ when Cor$(\xi_i,\xi_j)$ is high. Since every node makes most of its interactions within a cluster,  two nodes with highly correlated fluctuations $\xi(t)$ are likely to belong to the same community, and this can be enforced in the effective network by adding a connection between them}.

\subsection{Performance of the communities reconstruction} 
To quantify the effectiveness of  community reconstruction, we compute  $PE= \frac{m}{N}$, where $N$ is the total number of nodes and $m$ is the number of  nodes assigned to the wrong community.   We compute $PE$ for  $ \Delta\alpha$ between $0.05$ and $0.4$.  For each value of $\alpha$, we considered 50 different simulations by choosing different initial conditions.  The figure shows the plot of the mean of $PE$ and a shaded region corresponding to the standard deviation. For $ \Delta \alpha$ values larger than $0.4$, the reconstruction procedure cannot identify the communities correctly as synchronization rich club which appears.

\begin{figure}[h]
    \centering
    \includegraphics[width=1.0\columnwidth]{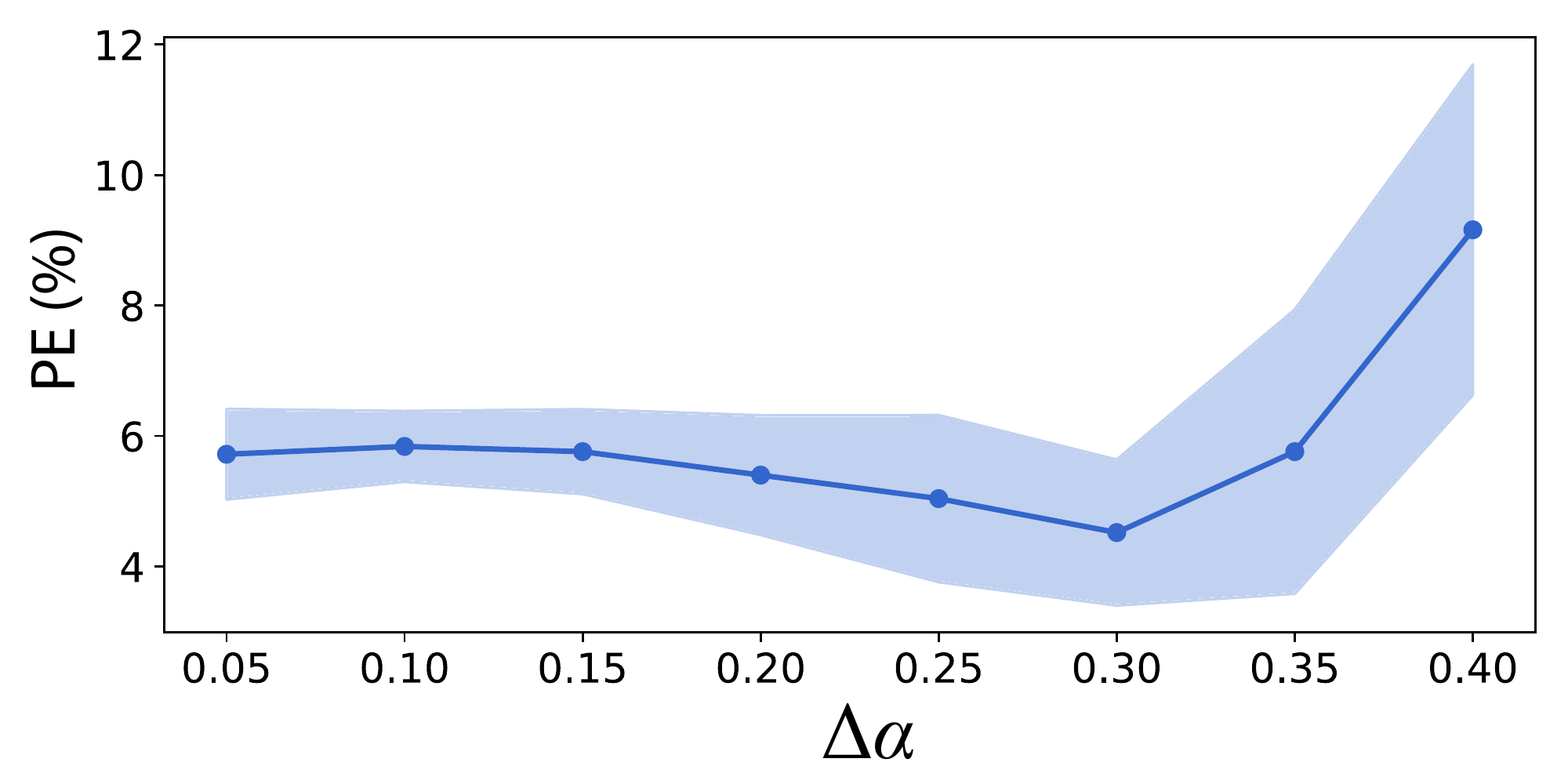}
\caption{{\bf Prediction error for misidentification of communities in the reconstructed cat cerebral cortex from synthetic data.} For each realisation, the chosen parameters are the same as in Figure \ref{RC2} and only the overall coupling is changed. Mean and standard deviation of prediction error (PE) computed for the network over 50 realizations for each value of $\alpha$. If $\Delta \alpha > 0.42$, the system synchronizes and the procedure cannot reconstruct the community structures.}
\label{fig:pe} 
\end{figure}

{ 
In the Appendix \ref{App:addsim}, we analyzed synthetic networks with 100 nodes which are undirected and have a rich-club structure. We used them as benchmark to evaluate the success of the reconstruction. The ability of the reconstruction procedure  to recover the community structure was  tested for various coupling functions and isolated dynamics. 
}

\section{Predicting critical transitions in rich-clubs}

The ability to reconstruct the network and dynamics from data can be exploited to predict critical transitions that may occur when the coupling strength varies. This is crucial for applications. For example in the cat brain, a transition to collective dynamics in the rich-club has drastic repercussions for the functionality of the network \cite{zamora-lopez2010,lopes2017}. 

The  goal is  to obtain and predict the onset of collective motion in the rich-club from data recorded when the network is far from a collective dynamics.   The effective network can predict the onset of such collective dynamics based on a single multivariate time series for fixed coupling strength in a regime far from the synchronized state.  We analyze time-series obtained simulating the dynamics  for $ \Delta\alpha = 0.3$, and reconstruct the network structure and the isolated dynamics.  

Transitions to synchronization between the scale variable is possible while the fast spikes remain out of synchrony \cite{rulkov2001}. Notice that the slow variable $w$ changes on a scale $1/\mu$. In the present setting we have $1/\mu = 10^3$ which  is about number of points we need to apply the approach. Thus, for such short time series we can neglect the slow scale. This is also an advantage of this present approach.  To estimate the transition to burst synchronization, we obtain the slow variable as a filter over the membrane potential (fast variable). Since we measure the membrane potential $y_i(t) = u_i(t)$, the slow variable is given as
$
z_i(t) = \mu \sum_{k=1}^t (y_i(k) - \sigma)
$
 and for a choice $\mu$ and $\sigma$ this can be identified with the slow variable of the model $w$.  {In Appendix \ref{app:predictions}, we derive the following equation for the slow variable of a node in the rich club:
 \[
 z(t+1) = (\lambda -  \Delta\alpha ) z(t) + \mu \sum_{n=0}^t z(n) 
 \] 
 where $\lambda=1.42$ is estimated from the data. The equation can be used to analyze the effect of the network connectivity on the dynamics. } We can use the data on the network and the dynamics recovered from the time-series recorded at  $\Delta \alpha = 0.3$  to predict that at the value $\Delta \alpha \approx 0.42$ the rich-club will develop a burst synchronization (details in Appendix \ref{app:predictions}).   %

\begin{figure}[t!]
        \centering
        \includegraphics[width=1.0\columnwidth]{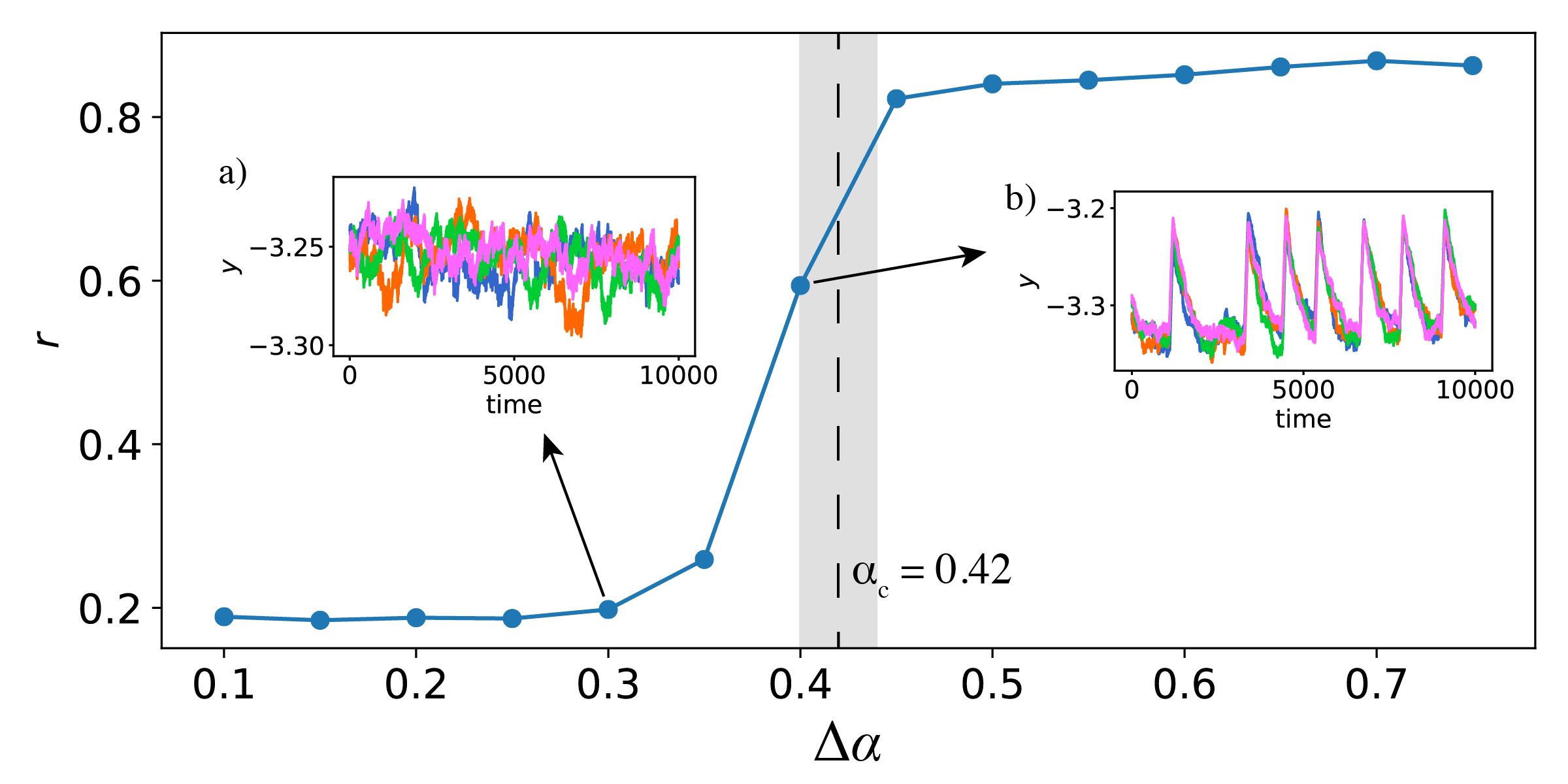}
\caption{{\bf Prediction of critical transitions in the rich-club of the cat cerebral cortex.} The level of synchronization $r$ of the rich-club is shown for different values of the coupling strength. Insets show time series of neuronal dynamics of four rich-club members and color of time series matches with the color of nodes in Fig.~\ref{RC2}. For values in the grey shaded region, $r$ is increasing towards close to one and the rich-club exhibits collective behavior. We can predict the critical coupling $\alpha_c$ (standard deviation in shaded region) by studying the effective network obtained from a time series measured at $\Delta \alpha=0.3$.}
\label{fig:criticality} 
\end{figure}

To capture a transition to a synchronized state, we introduce a phase $\theta_j(t)$ for the slow variable.  {To define $\theta_j(t)$, we first smooth the time series \cite{footnotesmoothing}. Then, we find the time $t_n$ of local maxima as the $n$th maximum point of the slow variable.  We introduce the phase variable $\theta$ as 
\begin{eqnarray*}
\theta_j(t) = 2\pi \left(\frac{t - t_n}{t_{n+1} - t_n} +t_n \right), \quad t_n < t < t_{n+1}
\end{eqnarray*}
as shown in Ref. \cite{pereira2007}. } We then compute the order parameter 
\begin{equation*}
r(t) e^{i \psi(t)} = \frac{1}{N_c} \sum_{j=1}^{N_c} e^{i \theta_j (t)}.
\end{equation*}
A small value of the order parameter, $r \approx 0$, means that no collective state is present, whereas $r(t)\approx 1$ means that the bursts are synchronized.   Figure \ref{fig:criticality} shows that behaviour of $r$ as a function of the coupling. The rich-club undergoes a transition to burst synchronization at $\Delta \alpha \approx 0.4$ that corresponds to an increase of roughly $40\%$ of the coupling strength and is close to the predicted value $\Delta \alpha \approx 0.42$.   In Appendix \ref{App:addsim}, we show other examples where the local dynamics is chaotic. 

\section{Obtaining  a statistical description of the network}

The effective network can provide statistical description of the network structure. To illustrate this, we reconstruct the statistical properties of scale-free networks.

\subsection{Scale-free networks of coupled bursting neurons} 

We consider coupled bursting neurons with excitatory synapses \cite{rulkov2001} in scale-free networks.   A scale-free network has degree distribution $P(k) =  C k^{-\gamma}$, where $\gamma>0$ is the characteristic exponent and $C$ is a normalising constant. We generate a scale-free network with $N=10^4$ nodes  such that the probability of having a node of degree $k$ is proportional to $k^{-\gamma}$, where $\gamma=2.53$.   We use a random network model which is an extension of the Erd\"os-R\'enyi  model for random graphs with a general degree distribution. More details are provided in Appendix D. 

\begin{figure}
    \centering           
        \includegraphics[width=1.0\columnwidth]{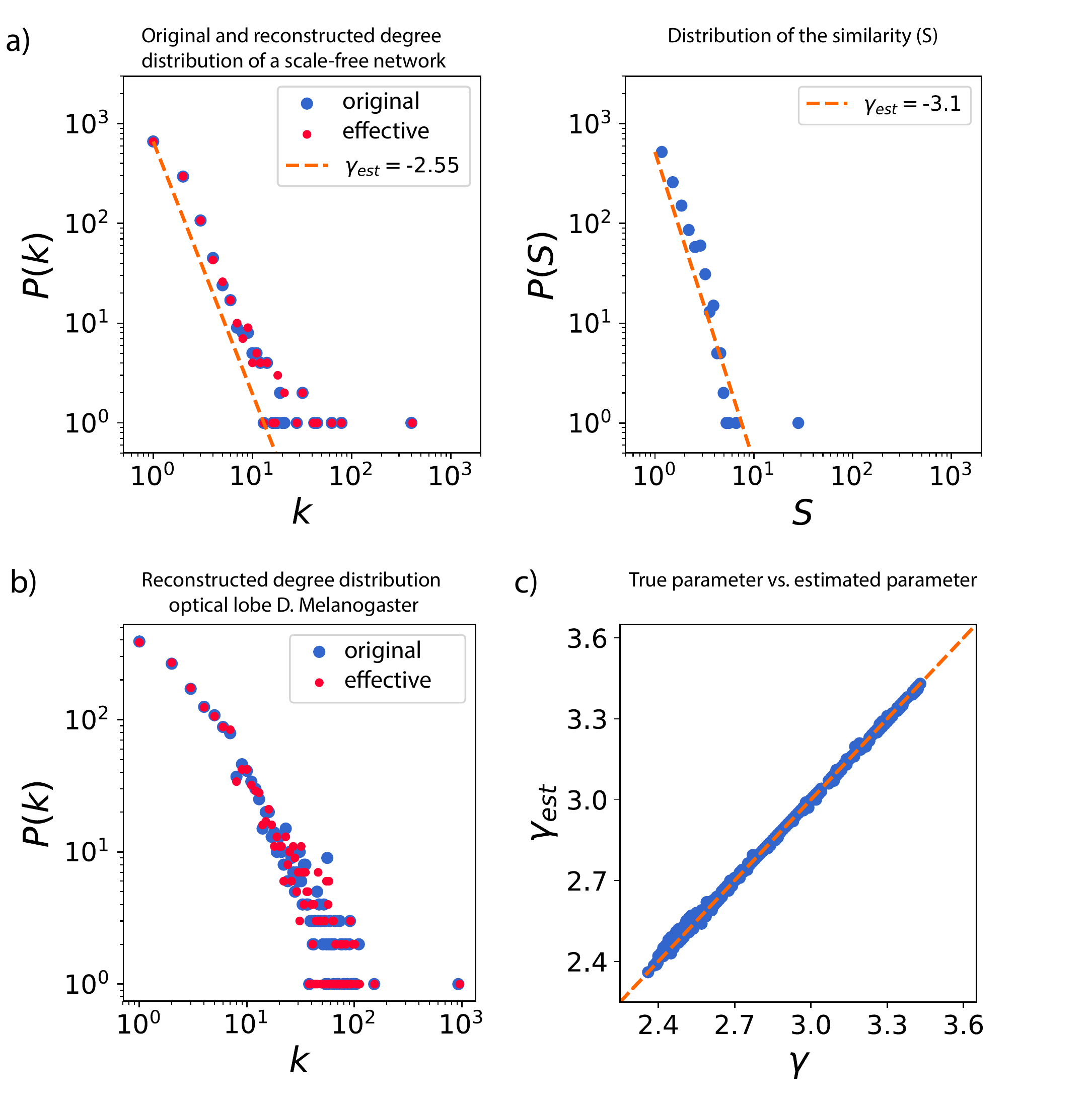}
\caption{ {\bf Reconstruction of  structural power-law exponents $\gamma$ of scale-free networks from data.}  We estimated $\gamma$ from the multivariate time series obtained from the dynamics random scale-free networks with degree distribution $P(k) \propto k^{-\gamma}$.  The  plots in panel (a) compare the functional and effective network approach.  We obtain better estimates using the effective network. Panel (b) shows the degree distribution of the original system (in blue) and that estimated from an effective model (in red) for the neural network in the optical lobe of Drosophila melanogaster. We obtained an accuracy of $3\%$ in the structural exponent $\gamma$. Panel (c) shows  the true exponent $\gamma$ versus $\gamma_{\rm {est}}$ obtained with an effective network from data for spiking neuron coupled with chemical synapses. We generated 1000 networks with distinct $\gamma$, from  which the $\gamma_{\rm est}$ estimate is within $2\%$ accuracy. }\label{fig2}
\end{figure}

For this reconstruction we only need 2000 data points for each node.  Again, to  every pair of  time series $y_i$ and  $y_j$ we assign  a \emph{Pearson distance} $s_{ij}\ge0$ and the node intensity  $S_{i} = \sum_{j}s_{ij}$. The empirical distribution of the intensities $S_i$ approximates the degree distribution of the network, see the second inset of Fig~\ref{fig2}(a). In the example here, the estimated structural exponent from the distribution of $S_i$ is $\gamma_{\rm est} = 3.1$, which yields a relative error of nearly 25\% with respect to the true value of $\gamma$ (see the plots in Figure \ref{fig2} a)). The functional network therefore overestimates $\gamma$, which has drastic consequences for the predicted character of the network. For example,  the number of connections of a hub for a scale-free network is  concentrated at $k_{\rm max} \sim N^{1/(\gamma - 1)}$, so the relative inaccuracy for the estimate $ k_{\rm est}$ of the maximal degree is $ k_{\rm max}/k_{\rm est}  = N^{1/\gamma - 1/  \gamma_{\rm est}}$, which is about  500\%. Such inaccuracy has important repercussions for the ability to predict the emergence of collective behaviour \cite{pereira2010,pereira2017}.  

The statistical measures used for the construction of a functional network typically depend in a nonlinear way on the degrees, thus causing a distortion in the statistics. We will discuss the case of Pearson distance. Suppose that the signals $\{(y_i(t),y_i(t+1))\}$  are purely deterministic, $y_i(t+1)=g_i(y_i(t))$. The Pearson distance $s_{ij}$ between the signal at $i$ and $j$ is a number between 0 and 1, depending on how close these graphs are. This distance depends nonlinearly on the degrees $k_i$ and $k_j$. Devising another distance $s'_{ij}$  without knowledge of the interaction, in general,  still carries the nonlinear dependence on the degrees. Once fluctuations from the network are included  the differences between time-series  can be due to fluctuations rather than differences in the degrees.  The decomposition of the rules in terms of interactions and fluctuations is essential to recover degree distribution accurately.

The effective network provides a  better statistical description of the network structure. To compare with the functional network approach, we constructed an effective network of the same system tested for the functional network. The estimate for $\gamma$ from the effective network is $\gamma_{\rm est}=2.55$, which has an error of only $1\%$ (inset one of Fig.~\ref{fig2}~(a)).  We repeat the analysis on a different network with different parameters $\gamma$ in the degree distribution. The estimated $ \gamma_{\rm est}$ values are shown in Fig.~\ref{fig2}~(c)  as a function of the true parameter $\gamma$. The relative error on the estimated exponent is within $2\%$. 

\subsection{Performance of the degree distribution reconstruction} 
In  Appendix \ref{App:addsim}, we present additional simulations
showing how accurate the degree distribution is reconstructed for various isolated dynamics. In particular in Figure 3 we show the results for a)  doubling maps with diffusive coupling,  b) logistic maps with Kuramoto interactions,  c) spiking neurons with electrical coupling, and  d) H\'enon maps with the y-component diffusive coupled with the x-component. Moreover, in Section III F we show the performance of the reconstruction for a system of differential equations coupled on scale-free networks.

We provide a study on the effects of noise in the reconstruction. We established that for stochastically stable \cite{Tanzi} systems such that the doubling map if the noise amplitude $\eta_0$ satisfies 
$\eta_0 < \alpha k_{\rm min}$, where $k_{\rm min}$ is the minimal degree the reconstruction procedure works. When the noise amplitude is of order $\alpha k_i$ nodes with degree less than $k_i$ cannot be  estimated.

\subsection{ The optic lobe of D. Melanogaster.} 

We applied our method to data simulated from the neuronal network in the Drosophila Melanogaster optic lobe, which constitutes $>$50\% of the total brain volume and contains 1781 nodes \cite{takemura2013}. The degree distribution has a power-law tail \cite{garcia-perez2018}. We used spiking neurons with chemical coupling to simulate the multivariate time series, from which we constructed an effective model  and estimate the degree distribution (Fig.~\ref{fig2}~(b)).

{
\subsection{Experimental data of optoelectronic oscillators}

We now apply our effective network to experimental data of networks of optoelectronic oscillators whose nonlinear component is a Mach-Zehnder intensity modulator. This data was generated in Ref. \cite{Hart2019} where the authors studied enhancement of synchronization by structural changes in the network. The the experimental setup can also be found in Refs. \cite{Hart2019,Hart2017}.  Each element consists of a clocked optoelectronic feedback loop. Light from a 780 nm continuous-wave laser is nonlinearly transformed as it passes through the Mach-Zehnder intensity modulator. Light intensity is converted into an electrical signal by a photoreceiver and measured by a field-programmable gate array (FPGA) via an analog-to-digital converter. The FPGA is clocked at 10 kHz, resulting in the discrete-time map dynamics of the oscillators. The FPGA controls a digital-to-analog converter  that drives the modulator with a voltage $x_i(t + 1) = \beta I (xi(t))$, closing the feedback loop. The elements are coupled  electronically on the FPGA according to the desired coupling matrix as described in detail in Ref. \cite{Hart2017}.  The system can be modeled as
\[
x_i(t+1) = \beta I(x_i(t)) +\sigma \sum_{j=1}^n A_{ij} [ I(x_j(t)) - I(x_i(t))]  \mbox{~mod~} 2\pi
\]
where $t$ is discrete time, $\beta$ is the feedback strength, $I(x) = \sin^2(x + \delta)$ is the normalized intensity output of the Mach-Zehnder modulator, $x$ represents the normalized voltage applied to the modulator, and $\delta$ is the operating point set to $\pi/4$.   The data is acquired for $\beta = 4.5$ and $17$ elements coupled through the network  presented in Figure \ref{Expt} left panel.  The coupling strength $\sigma$ varies from $0$ to $1$ in steps of $0.0325$ starting from $0.015625$. For each fixed value of $\sigma$, we obtain the experimental multivariate time series $\{ x_{1}(t), \cdots, x_{17}(t)\}_{t=1}^{15385}$. 

We discard the first $5000$ data points for each $i={1,\cdots, 17}$ as a transient. We will provide an analysis for the coupling $\sigma = 0.03125$. First, we perform a functional network analysis by considering a correlation matrix $\Sigma_{x}$ of the multivariate time series. To obtain a model of the adjacency matrix we threshold $\Sigma_{x}$. The value of the threshold $0.02$ is chosen such that the functional network has a mean degree close to the actual network. The result is shown in Figure \ref{Expt} in the middle panel and as observed the functional network does not capture the actual network structure. 

Next we employ the effective network. We start by applying Step (1) to learn the function $g_i$ and Step (2) from where we obtain the degrees and coupling strength.  Once we obtain $g_i$, we filter the determinist part from $x_i$ to obtain the fluctuations $\xi_i$. Next, we compute  correlation matrix $\Sigma_{\xi}$ for the fluctuations $\xi_i$. To turn this matrix into a network, we threshold it. Again the value of the threshold is fixed such that the mean degree is closed to the actual network. Here, any threshold value from $0.07$ to $0.1$ works. The result is shown in Figure \ref{Expt}, in the right panel, and shows excellent agreement with the actual network. In fact, only two links are misidentified. 

We also performed the analysis for further coupling strengths $\sigma$. For large coupling strengths both functional network and effective network will capture the network misidentifying on average 4 links. In these cases, the effective network has the advantage that it provides in addition to a model of the adjacency matrix also a model for the local dynamics.

\begin{figure*}[t!]
\centering
\includegraphics[width=.8\textwidth]{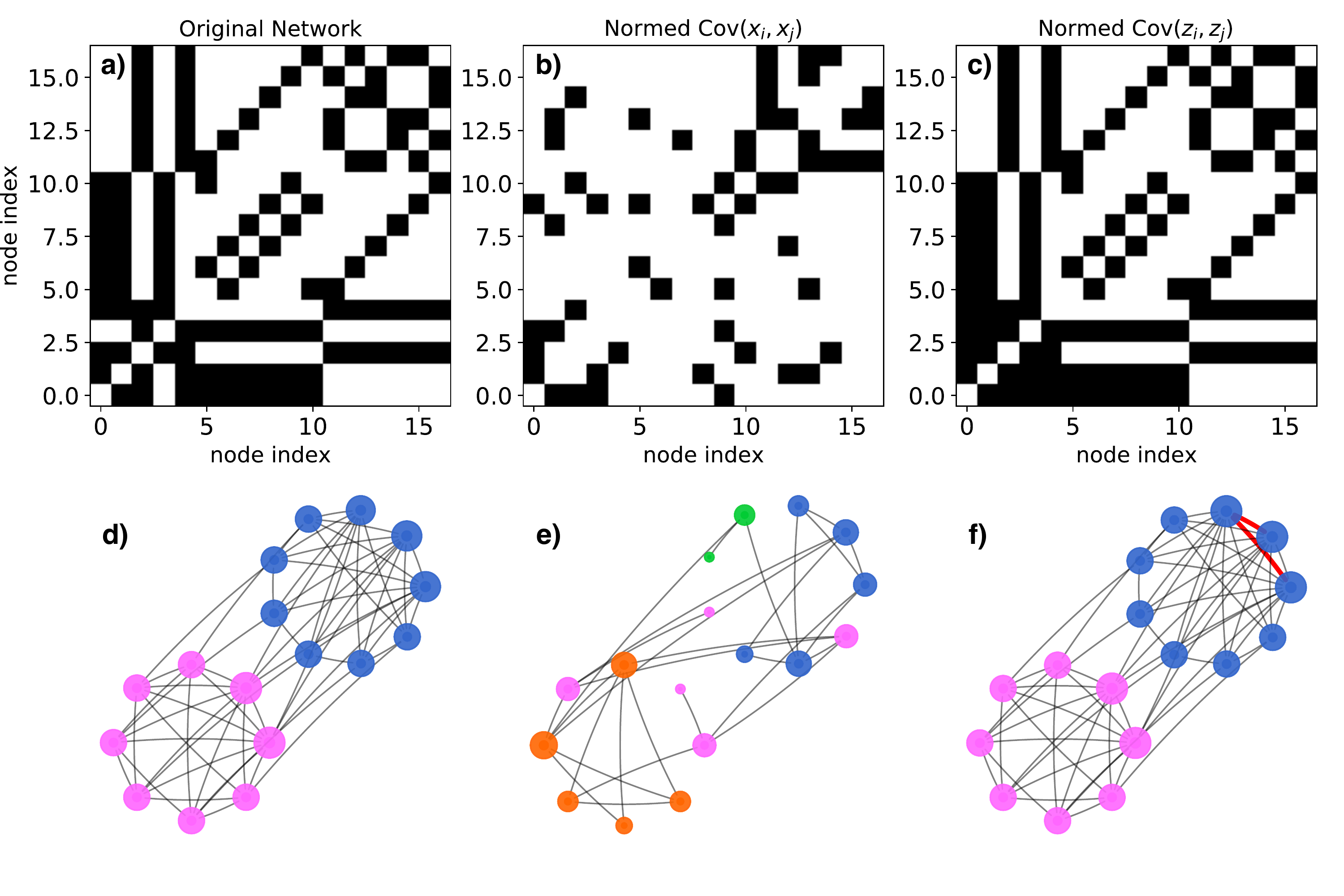}
\caption{
{
{\bf Effective network from experimental data of networks of optoelectronic oscillators.}  
We consider multivariate time series of voltages of a network of $17$ weakly coupled optoelectronic oscillators (interaction corresponding to $1\%$ of the oscillator amplitude).
In the left panel, we show the actual network used to coupled the optoelectronic oscillators in Ref.  \cite{Hart2019} as adjacency matrix in a) and graph representation in d). In the mid panel, we show the  reconstruction of the network by a functional network analysis in terms of its adjacency matrix in b) and graph representation in e). In the right panel, we show the reconstruction of the network from an analysis of the dynamical fluctuations by applying the effective network approach as adjacency matrix in c) and graph representation in f). The effective network provide a striking reconstruction and only two links are misidentified and are indicated in f as red links. In the graph representation, the nodes of the network are coloured according to the community obtained by a community detection algorithm \cite{blondel2008}.}}
\label{Expt} 
\end{figure*}

}

\section{Conclusions}
We have introduced an {\em effective network} obtained from time-series of a complex network observing the dynamics at each node. Our method complements the existing ones in two ways: First of all, it encompasses the case of  chaotic local dynamics at each node. Secondly it deals with weak coupling among the nodes. Both cases are commonly found in applications \cite{kandel2000,schneidman2006,haas2015}. Key to the success of the reconstruction is the heterogeneity of the network which allows us to perform a multi-level reduction. 
To recover the community structures, we use that  certain noise terms associated with the time series at two nodes in the same community are correlated. By collecting data when the network is far from critical transitions, an effective network  enables us to predict a critical transition.

We have compared our procedure with methodologies  most relevant for the systems  considered. We have  excluded results tailored to specific setups or dynamics (binary dynamics \cite{li2017universal}, and see \cite{wang2016} for a review). We did not consider methods that rely on measurements  obtained by intervening on the system with controlled inputs \cite{nitzan2017revealing} and restrict our attention to time-series  recorded under constant conditions.  When the coupling is strong, sparse recovery can be applied \cite{brunton2015}. When the coupling is weak sparse recovery cannot distinguish small parameters from those that are identically zero thus misidentifying connections between nodes. 
 Also model-free methods  are ill-suited as the influence of a single pairwise interaction on the time-series is  weak and can hardly be detected.

 {The effective network methodology performs well when the network is heterogenous and has a few nodes making a large number of connections while most of the nodes are less connected, and the local dynamics are chaotic and their typical orbits visit most of the phase space. The effective network approach did not perform well in two cases. The first is when most of the observed time-series take values on a very restricted part of the phase space, for example if the local dynamics has a singular attractor, as an attracting fixed point, or if it spends long periods of time in a small region, like around the fixed points of the classical Lorenz attractor. This means that we don't have access to a big portion of phase space,  and no prediction is possible in those regimes of coupling strength that make these portions accessible. The passage  near a fixed point also suppresses the fluctuations  hindering the reconstruction of communities. This is what seems to happen for example in the bursting dynamics of Rulkov maps, when the quiescent state is too long. These situations are excluded if the local dynamics is sufficiently chaotic. The second case is when the coupling  is strong enough to synchronize big parts of the network. For example, a synchronous rich-club can send similar forcing to nodes in different communities resulting in high correlations between the fluctuations. Therefore our method would identify these nodes as belonging to the same community even if they are not. }
%
%\tpa{}

\appendix
{
\section{Effective network representation from data}
\label{app:en}
A summary of the effective network approach is given in Figure \ref{RC}. Here we include some details that were omitted for the sake of presentation in the main text. 

 \begin{figure*}[t!]
        \centering
        \includegraphics[width=\linewidth]{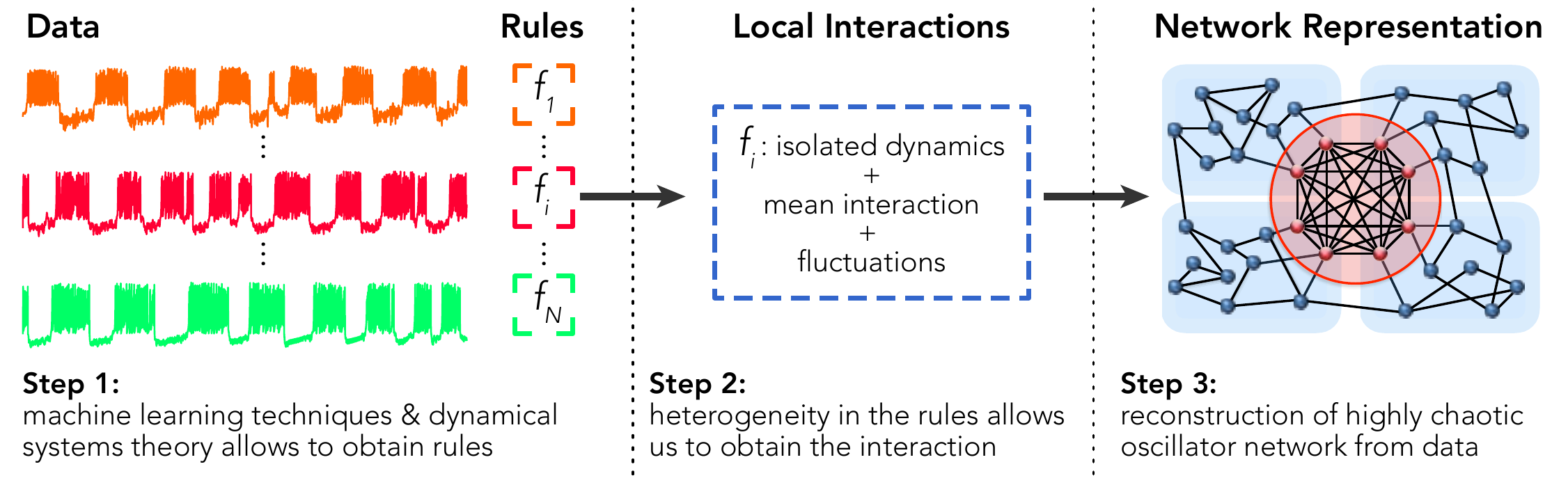}
       % \caption{the degree distribution}
\caption{Reconstruction scheme with the effective network.  From the time series, we build a model for the local evolution  $f_i$ at each node. Under the assumption that such rules change from node to node depending on their connectivity, we estimate the coupling function. Using the fluctuations of the time series with respect to the low-dimensional rules, we recover the community structures. Gathering all this information, we obtain an effective network  that can be used to predict critical transitions}.
\label{RC} 
\end{figure*}

In Step 2 of the reconstructing procedure, we identify low degree nodes by analysing the distribution of $S_i$. More precisely, we use the top $N_{\rm top}$ nodes of the highest intensity to obtain a proxy for the isolated dynamics. We then average these rules to get $\langle g \rangle \approx f$. The choice of $N_{\rm top}$ is not fixed and depends on the number of nodes and the fluctuation  $\sigma_g^2 = \langle (g_i - \langle g \rangle)^2 \rangle$.  { For scale-free  (Barabasi-Albert)  networks the degree of the hubs scales as $N^{1/2}$, a good heuristic
is to choose $N_{\rm top}$ satisfying $\sigma_g^2 / N^{1/2}_{\rm top} \ll 1.$}
The effective coupling function  $\alpha \kapppa_i  v$ can be obtained analysing the family $\{ g_i - \langle g \rangle \}_{i=1}^{N}$ which can yield the shape of $v$ up to a multiplicative constant via a nonlinear regression by imposing that $g_i - \langle g \rangle$ and $g_j - \langle g \rangle$ are linearly dependent. The choice of the base function for the fitting is supervised (see Appendix \ref{App:addsim}).

In Step 3, after selecting a $v$ that satisfactorily approximates $g_i- \langle g \rangle$ up to a multiplicative constant over all indices $i$,  the parameter $\beta_i$ is estimated  using a dynamic Bayesian inference. Because the fluctuations $\xi_i(t)$ are close to Gaussian, we use a Gaussian likelihood function and a Gaussian prior for the distribution of the values of $\beta_i$, and hence obtain equations for the mean and variance. We split the data into epochs of $200$ points and update the mean and variance iteratively.   
}

%\noindent
\subsection{Community structures} 

Once we obtain the rules $g_i$, we filter the deterministic part of the time series $y_i$ and access the fluctuations $\xi_i$ (recall Eq. \eqref{eq:meanfield}) and decompose it as  $\xi_i = \xi_i^{\rm c} +  \xi_i^{\rm o}  $ where $\xi_i^{\rm c}$ is the fluctuation of the local mean field from nodes in the cluster containing  $i$, and  $\xi_i^{\rm o}$ is the contribution from outside the cluster. Since a node makes most of its connections within its cluster, $\xi_i^c\gg\xi_i^{o}$ with high probability,  and thus if $i$ and $j$ belong to the same cluster $ \mbox{Corr}(\xi_i , \xi_j)=\mbox{Corr}(\xi^c_i , \xi^c_j)$. The common noise is generated by the common connections between nodes $i$ and $j$. For fixed isolated dynamics and coupling function 
\begin{eqnarray*}
\mbox{Corr}(\xi^c_i , \xi^c_j) \propto \widehat{\mu}_{ij}.
\end{eqnarray*}
\noindent
$\mbox{Corr}(\xi_i , \xi_j)$ is related to the \emph{matching index}~\cite{zamora-lopez2010} of the nodes $i$ and $j$. This is a parameter used to quantify the number of common neighbours of two nodes.  Recall that the degree of node $i$ is $ k_{i} = \sum_{j}^N A_{ij}$ and counts the number of neighbours it has. Consider the neighbourhood of node $i$, $\Gamma (i) = \{j \in \{ 1, \dots, N\} \, | \,  A_{ij} = 1\}$. This is the set of nodes that shares an edge with the node $i$. The matching index of nodes $i$ and $\ell$ is the cardinality of the overlap of their neighbourhoods  $\mu_{i \ell} = |\Gamma(i) \cap \Gamma(\ell)|$. We consider the normalised matching index:
\begin{eqnarray*}
\widehat{\mu}_{i\ell} = \frac{|\Gamma(i) \cap \Gamma(\ell)|}{|\Gamma(i) \cup \Gamma(\ell)|} 
\end{eqnarray*}
or equivalently in terms of the adjacency matrix
\begin{eqnarray*}
\widehat{\mu}_{i \ell} 
= \frac{(A + A^2)_{i \ell}}{k_{i} + k_{\ell} - (A + A^2)_{i\ell}}.
\end{eqnarray*}
Clearly $\widehat{\mu}_{i\ell} = 1$ if and only if $i$ and $l$ are connected to exactly the same nodes,and $\widehat{\mu}_{i \ell} = 0$ if they have no common neighbours. 
It is well known that in the cat cerebral cortex nodes in the same community have a high matching index while nodes are distinct communities has a low matching index. This tends to be typically in modular networks \cite{zamora-lopez2010}. For nodes in distinct clusters the component $\xi_i^{\rm c} \approx 0$, so $ \mbox{Corr}(\xi_i , \xi_j) \approx 0.$ We recover the network structure from a {\em noise covariance analysis}.

%\subsection{Filtering out the deterministic part} 

Filtering out the deterministic part plays a major role in recovering community structures. Suppose we have two signals of the form $y_i(t) = {Y}_i(t) + \zeta(t), \quad i=1,2 $,  where ${Y}_i$ is independent  of $i$ and $\zeta(t)$ is a common noise term. $Y_i$ represents the superposition of the deterministic chaos and the independent fluctuations. For the correlation, we have
\begin{eqnarray*}
\mbox{Corr}(y_i , y_j) \approx \frac{\mbox{Cov}(\zeta,\zeta)}{\sigma^2_{Y_1} \sigma_{Y_2}^2}
\end{eqnarray*}
\noindent
Hence, the large values of the variance of the time series ($\sigma_{y_i}\approx \sigma_{Y_i}\gg \sigma_{\zeta}$) suppress the contribution of the common noise, and an analysis solely based on the the original time series $y_i$ will overlook the common noise contribution.

\section{Functional networks} 
\label{app:func_networks}

For networks of chaotic oscillators, building the functional network from the  standard Pearson correlation between time series gives no meaningful results because of the  decay of correlation intrinsic to dynamics. Functional networks are built using a Pearson distance $s_{ij}\ge 0$  describing the proximity of the dynamics at two nodes $i$ and $j$.  To do this,  we consider the time series  ${z}_i(t):=({y}_i(t),{y}_i(t+1))$, $t=0,\dots,T-1$ reordered in $z_i^{\rm lex} (t)$ according to the lexicon order; that is, according to the magnitude of the first component of  $z_i(t)$. Then, let $r_{ij}$ be the Pearson correlation,  $r_{ij}=$ Cor$(z_i^{\rm lex}, z_j^{\rm lex})$, so  that $r_{ij}=1$ indicates that the attractors at nodes $i$ and $j$  agree.  Define the Pearson distance $s_{ij}=1-|r_{ij}|$ so that $s_{ij}=0$ indicates  agreement of the dynamics and $s_{ij}>0$ measures the difference between the attractors.

The intensity $S_i =  \sum_{j}s_{ij}$  approximates how many nodes have a dynamical rule different from $i$ and helps to distinguish between poorly connected nodes and hubs. Since  most of the network is composed of poorly connected nodes, they exhibit a smaller $S_i$ than high-degree nodes, which are scarcer and have different dynamics  from the low-degree nodes.

\section{Predicting critical transitions}
\label{app:predictions}
Here we explain how to gather the information for a theoretical prediction of the critical transition.

\noindent
{\bf Reduction in the rich-club}. Nodes in the rich-club have degrees of approximately $\Delta$ and make $\kappa \Delta$ connections inside the rich-club and $(1-\kappa)\Delta$ connections to the rest of the  network. Following our reduction scheme, the interactions within and outside the rich-club  can be described by the expected value of the interactions with respect to the invariant measure associated with each of them.  Let $C$ denote the set of nodes in the rich-club, then the coupling term is
\begin{eqnarray*}
\sum_{j} A_{ij} \bm{H}(\bm{x}_i,\bm{x}_j) =  \sum_{j \in C} A_{ij} \bm{H}(\bm{x}_i,\bm{x}_j) + \sum_{j \not \in C} A_{ij} \bm{H}(\bm{x}_i,\bm{x}_j)
\end{eqnarray*}
However,
\begin{eqnarray*}
\sum_{j \not \in C} A_{ij} \bm{H}(\bm{x}_i,\bm{x}_j) = (1 - \kappa)\Delta \int \bm{h}(\bm{x}_i, \bm{y}) d\mu(\bm{y}) + \bm{\xi}^o_i(t)
\end{eqnarray*}
where $\mu$ is the invariant measure for the nodes outside the rich-club. Hence,  for the rich-club we obtain 
\begin{eqnarray*}\label{RCeq}
\bm{x}_i(t+1) = \bm{q}_i(x_i(t)) +  \sum_{j \in C} A_{ij} \bm{H}(\bm{x}_i,\bm{x}_j) + \bm{\xi}^o_i(t),
\end{eqnarray*}
where 
\[
\bm{q}_i(\bm{x}_i(t)) = \bm{F}_i(\bm{x}_i(t)) + (1 - \kappa)\Delta\alpha \int \bm{H}(\bm{x}_i, \bm{y}) d\mu(\bm{y}).
\]
{\bf Predicting the transition to collective behaviour.}
Let us recall that when isolated 
$
u_i(t+1) = F_{1,i}(u_i(t))+ w_i(t)$ where $F_{1,i} \approx F_1$, $ w_i (t+1) = w_i (t) + \mu (w_i(t) - 1)$,  and
\begin{eqnarray}
\label{yeq}
w(t+1) = w_0 + \mu \sum_{n=0}^t (u(n) - 1))
\end{eqnarray}
Using the reduction Eq. (\ref{RCeq}), in the network we obtain
\begin{eqnarray*}
u_i(t+1) = F_{1,i}(u_i(t)) + u_i(t) +  \Delta\alpha [ \langle u \rangle - u_i(t)] + \xi_i(t)
\end{eqnarray*}
where $i$ denotes the $i$th nodes in the rich-club, $\langle u \rangle$ is the mean  in the rich-club and $\xi_i$ are fluctuations. We fix two nodes  $y_i = u_i$ and $y_j = u_j$ in the rich-club and consider
\begin{eqnarray*}
\zeta(t) = u_i(t) - u_j(t)
\end{eqnarray*}
Using that $F_{1,i} \approx F_1$ by the mean value theorem we obtain
\begin{eqnarray*}
\zeta(t+1) = DF_1 (x_i(t)) \zeta(t) + \mu \sum_{n=0}^t \zeta(n) -  \Delta\alpha \zeta(t)
\end{eqnarray*}
and introducing a proxy for the dynamics of the slow variables  
\begin{eqnarray*}\label{ueq}
\eta(t) = \sum_{n=0}^t z(n)
\end{eqnarray*}
and considering 
$\sum_{n=0}^t DF_1(x_i(n)) \zeta(n) \approx \lambda \sum_{n=0}^t \zeta(n)
$ where we used that $\sum_{n=0}^t \zeta(n)$ is a slow variable.   We obtain
\begin{eqnarray*}\label{udyn}
\eta(t+1) = (\lambda -  \Delta\alpha ) \eta(t) + \mu \sum_{n=0}^t \eta(n)  
\end{eqnarray*}
For the cat cerebral cortex $\Delta = 37$. Given the time series $\{y_i \}$ for $ \Delta\alpha = 0.3$, we estimate $F_1$ using our method $i$ as the slow variables are constants over short time scales, and the obtain slow variables as a filter over the fast variables.   From the data,  we estimate
$
\lambda = 1.42
$
and thus we obtain 
$
 \Delta\alpha = 0.42.
$
At this critical value the slow variables tend the stay together due to the contraction in the dynamics. This is related to the onset of synchronization in the bursts, which is captured via a phase variable through the order parameter.

%%%%%%%%%%%%%%%%%%%%%%%%%%%
For  estimation of the power-law distribution parameters, we use the maximum likelihood estimator  \cite{muniruzzaman1957,hill1975}. After that, we test the reliability between the data and the power law by using the goodness-of-fit method. If the resulting $p$-value is larger than 0.1, the power-law estimation is an appropriate hypothesis for the data. A complete procedure for the analysis of power-law data can be found in Ref.~\cite{clauset2009}. 

{
\section{Dimensional reduction in heterogeneous networks}
We  present an informal statement of the theoretical results used in the reconstruction procedure. For a precise statement  see \cite{pereira2017}. The theorem has three main assumptions:

\begin{itemize}
\item[1)] {\bf The local dynamics must increase the distance between points} by a constant factor.\looseness=-9
\item[2)]{\bf The networks are heterogeneous}. Most of the nodes have small degree $\delta\sim N^{\frac{\epsilon}{2}}$, and some nodes are hubs with degree $\Delta\sim N^{\frac{1}{2}+\epsilon}$.
\item[3)] {\bf The reduced dynamics must be hyperbolic}. The maps $\bo G_j$ are either expanding or to have a finite number of attracting periodic orbits. In dimension one, every map can be perturbed by an arbitrarily small amount to obtain such an hyperbolic map \cite{Strien}.
\end{itemize}
Under these assumptions, we have the following result

\begin{theorem}[\cite{pereira2017}]
For every hub node $j$, the dynamics at the hub is given by
\begin{eqnarray*}
\bo x_j(t+1)=\bo G_j(\bo x_j(t))+\bo \xi_j(t)
\end{eqnarray*}

where $|\bo\xi_j(t)|<\xi$ for time $T$ with $1\leq T\leq \exp[C\xi^2\Delta]$, and a set of initial condition of measure $1-T/\exp[C\xi^2\Delta]$, where $C$ is constant in $\Delta$ and $\xi$.
\end{theorem}

Notice that one can pick the time scale $T$ exponentially large, but such that $T/\exp[C\xi^2\Delta]$ is very small so that, for large $\Delta$, the approximation result holds for very long time and for a large set of initial conditions.}

{
\section{The effective network for a variety of chaotic dynamics and coupling}\label{App:addsim}
We tested the  performance of the effective network in recovering community structure and degree distribution for the systems listed below. Recovery of community structures was tested on a network of $100$ nodes having five clusters of $20$ nodes each. Four of these clusters are modeled as Erd\"os-Renyi random graph with  connection probability $p=0.3$, and the fifth, the integrating cluster,  with $p=0.8$. The coupling strength is $\alpha$ is of the order of $10^{-4}$.  Recovery of degree distribution was tested on scale-free networks with 6000 nodes and characteristic exponent $\gamma$ varying between 2.4 and 3.6, and coupling strength at $\alpha \Delta=0.5$. Details and results of the simulations can be found in Supplementary Materials.
%\begin{enumerate}

{\bf Doubling maps.}
Since the dynamics is one dimensional, we denote $\bm{x} = x$ and $\bm{F}_i(\bm{x}) = f_i(x)$ with  
%\begin{eqnarray*}
$
f_i(x) = 2x + \varepsilon_i \sin 2\pi x \mod 1$
%\end{eqnarray*} 
and where we take $\varepsilon_i$ to be i.i.d. random variables uniformly distributed on $[0,10^{-3}]$. Likewise we write $\bm{H} = h$ with 
%\begin{eqnarray*}\label{Eq:DiffCoup}
$
h(x_j,x_i) = \sin2\pi x_j-\sin2\pi x_i.
$
%\end{eqnarray*}
%We fixed $\alpha = 10^{-2}.$
%
We were able to recover all community structures, and the characteristic exponent $\gamma$ within 0.5\% accuracy.

{\bf Logistic map.} Again, $\bm{x} = x$ and $\bm{F}_i(\bm{x}) = f_i(x)$ where  
%\begin{eqnarray}
$
f(x) := 4 x(1-x),
$
%\end{eqnarray}
and we consider 
%\begin{eqnarray}
$
h(x_j,x_i) = \sin(2\pi x_j-2\pi x_i).
$
%\end{eqnarray}
%
We were able to recover all community structures, and the characteristic exponent within 0.5\% accuracy.
 
{\bf Spiking neurons with electrical synapses. }We use the same  spiking neurons as in the main body of the manuscript and  denoting $\bm{x} = (u,w)$ the coupling function  reads as
%\begin{eqnarray}\label{Eq:ChemCoup}
$
\bm{H}(\bm{x_i},\bm{x_j})=\bm{E} ( \bm{x}_j - \bm{x}_i  ) = (u_j - u_i,0).
$
%\end{eqnarray}
We were able to recover all community structures and the characteristic exponent within 2\% accuracy.

{\bf Bursting neurons with electrical synapses.} Our  numerical investigation reveals that when the resting time is not much larger then the total bursting time the reduced dynamics is capable of extracting the relevant information of the time series. Thus, we fixed the neuron parameter $\beta = 4.4$ to obtain a bursting dynamics.  The coupling is electrical as for the systems above.
We were able to recover all community structures.

{\bf Henon Maps.} Using the notation $\bm{x} = (u,w)$, the coupled H\'enon maps we study are given by
%{\small
%\begin{eqnarray}
$    \bm{F}(u,w)=
%\left\{
%                \begin{array}{ll}
(1-1.4u^2 + w, 0.3 w$
%                \end{array}
%              \right.   \mbox{~ and ~ }
and
$
\bm{H}(\bm{x_i},\bm{x_j}) = 
%\left\{
%                \begin{array}{ll}
( w_j-w_i,  0)$.
%                \end{array}
%              \right.
%\end{eqnarray}
%}
%
We assume to observe only the dynamics of the first component  $ y = \phi(\bm{x}) = u$. 
In this multidimensional case, the reconstruction will start by determining the dimension  
of the reduced system. Takens embedding reveals that the dimension is two for large time excursions,  hence, we will aim at learning a function

\begin{eqnarray}
 y_i(t+1) = g_i(y_i(t),y_i(t-1)) + \xi_i(t).
\end{eqnarray}
We use polynomial functions for the fitting via a 10-fold cross-validation.  
Our theory implies that 
$g_i(y_i(t),y_i(t-1))  = f (y_i(t),y_i(t-1)) + \alpha k_i v(y_i(t),y_i(t-1)) 
$ where $f$ models the isolated dynamics and $v$ the coupling. We obtain $f$ from the low-degree nodes via a similarity analysis. We learn $h$ by 
$
\alpha k_i v(y_i(t),y_i(t-1))  = g_i(y_i(t),y_i(t-1))  - f (y_i(t),y_i(t-1)).
$
We were able to recover all community structures and the characteristic exponent within 2\% accuracy.
 }

\subsection*{Data Archival}

The connection matrices of cat cortex is found at {\small \url{https://sites.google.com/site/bctnet/datasets}}. Connectivity of Drosophila Melanogaster is found at {\small \url{https://neurodata.io/project/connectomes/}. The experimental data on the optoelectronic oscillators from Ref.  \cite{Hart2019} can be obtained by contacting Joseph Hart and R. Roy upon reasonable request.}

%\bibliography{scibib}
%
%\bibliographystyle{Science}

\section*{Acknowledgments}
We are in debt with Joseph Hart and Raj Roy for sharing the experimental data with us.  We thank Tomislav Stankovski, Chiranjit Mitra, Mauro Copelli, Dmitry Turaev and Jeroen Lamb for enlightening discussions. This work was supported in part by FAPESP Cemeai grant 2013/07375-0, the European Research Council (ERC AdG grant number 339523 RGDD), TUBITAK Grant No. 118C236 and the Serrapilheira Institute.

\end{document}